# Exploration of Whether Skylight Polarization Patterns Contain Three-dimensional Attitude Information

Huaju Liang,[1] Hongyang Bai,[1,*] and Tong Zhou[2]

[1]*School of Energy and Power Engineering, Nanjing University of Science and Technology, Nanjing 210094, China*
[2]*School of Mechanical Engineering, Nanjing University of Science and Technology, Nanjing 210094, China*
*\*hongyang@mail.njust.edu.cn*

**Abstract:** Our previous work has demonstrated that Rayleigh model, which is widely used in polarized skylight navigation to describe skylight polarization patterns, does not contain three-dimensional (3D) attitude information [1]. However, it is still necessary to further explore whether the skylight polarization patterns contain 3D attitude information. So, in this paper, a social spider optimization (SSO) method is proposed to estimate three Euler angles, which considers the difference of each pixel among polarization images based on template matching (TM) to make full use of the captured polarization information. In addition, to explore this problem, we not only use angle of polarization (AOP) and degree of polarization (DOP) information, but also the light intensity (LI) information. So, a sky model is established, which combines Berry model and Hosek model to fully describe AOP, DOP, and LI information in the sky, and considers the influence of four neutral points, ground albedo, atmospheric turbidity, and wavelength. The results of simulation show that the SSO algorithm can estimate 3D attitude and the established sky model contains 3D attitude information. However, when there are measurement noise or model error, the accuracy of 3D attitude estimation drops significantly. Especially in field experiment, it is very difficult to estimate 3D attitude. Finally, the results are discussed in detail.

## 1. Introduction

Many insects can extract orientation information by processing the polarized light in the sky, which provides a new idea for the navigation of guided projectiles, vehicle and so on [2], such as the orientational cue is captured by desert ants from the sky to follow a straight line to migrating across continents [3, 4]; Locusts transform polarized light information as a compass [5, 6]; Monarch butterflies perceive the polarization of skylight to determine orientation [7, 8]. Insects sense polarized skylight as a compass through a special region of their compound eyes named as dorsal rim area (DRA) [9-11]. And the reason why skylight can provide a directional reference is that skylight has polarization patterns [12-14], which are caused by the scattering of sunlight from the atmosphere and usually described by Rayleigh model.

Rayleigh model is based on Rayleigh scattering theory, and can well explain why the sky is blue [15, 16]. This model is convenient and easy, so it is widely used in polarized skylight navigation. And based on this model many one-dimensional attitude estimation methods have been proposed, which can estimate orientation, such as Zenith method [12, 17-22], Symmetry Detection method [23, 24], solar meridian and anti-solar meridian (SM-ASM) method [25, 26]. In addition, some two-dimensional attitude estimation methods have been proposed [27-29], which can extract solar vector to determine two Euler angles.

However, with the development of society and defense, there are more and more demands for three-dimensional (3D) attitude estimation. So, several methods have been proposed to estimate 3D attitude. These methods includes integrated navigation methods [30-33] and unreal-time methods based only on Rayleigh model [34]. Moreover, several real-time methods based only on Rayleigh model have been proposed to estimate 3D attitude in real-time [35-39]. However, our previous work has demonstrated that Rayleigh model does not contain three-dimensional (3D) attitude and thus 3D attitude cannot be estimated in real-time based only on Rayleigh model [1]. Rayleigh model only considers single Rayleigh scattering event [15, 40, 41], which cannot fully describe the information of the skylight, so it is still necessary to further explore whether the skylight polarization patterns and other sky models contain 3D attitude information.

To explore whether the skylight polarization patterns contain the information of 3D attitude, based on the social spider optimization (SSO) [42-44], a SSO algorithm is proposed to estimate three Euler angles in this paper. This algorithm considers the difference of each pixel among polarization images based on template matching (TM) [45], so it can make full use of the polarization information captured by polarization imager. According to the observation, there are four neutral points in the sky [46], which cannot be described by Rayleigh model. So, Berry model has been proposed to describe the angle of polarization (AOP) and degree of polarization (DOP) of skylight based on four neutral points [47-51]. However, Berry model cannot describe the light intensity (LI) information in the sky, which can be well described by Hosek model [52-54]. So, this paper established a combines Berry model with Hosek model to fully describe the AOP, DOP and LI information of skylight. In addition, the established sky model considers the influence of ground albedo, atmospheric turbidity, and wavelength of incident light [49, 51].

In short, to explore whether the skylight polarization patterns contain 3D attitude information a SSO method based on TM is proposed to estimate three Euler angles. And a sky model is established based on Berry and Hosek model, which considers the influence of four neutral points, ground albedo, atmospheric turbidity, wavelength of light. Finally, simulation and experiment are carried out and the results are discussed.

## 2. Sky model

To explore whether the skylight polarization patterns contain the information of 3D attitude, an accurate sky model must be established. So, in this section, a sky model is established based on Berry model and Hosek model to fully describe the AOP, DOP and LI information in the sky. And the established sky model considers the influence of neutral points, ground albedo, atmospheric turbidity, wavelength of light.

AOP:

According to four neutral points in the sky, Berry model uses complex numbers to describe skylight polarization patterns [47, 48, 50].

$$w(\zeta) = -\frac{4(\zeta - \zeta_+)(\zeta - \zeta_-)(\zeta + 1/\zeta_+^*)(\zeta + 1/\zeta_-^*)}{(1+|\zeta|^2)^2 |\zeta_+ + 1/\zeta_+^*| |\zeta_- + 1/\zeta_-^*|} \tag{1}$$

where $\zeta_+$, $\zeta_-$, $-1/\zeta_+^*$ and $-1/\zeta_-^*$ are four complex numbers and represent the positions of four neutral points in the sky, which are influenced by solar zenith, solar azimuth, atmospheric turbidity, wavelength of light. $\zeta$ is a complex number and represents the projection of the observed point on the complex plane, as shown in Fig. 1. $|\cdot|$ represents the 2-norm of a complex number. Then, the AOP of skylight is given by

$$AOP(\zeta) = \arg[w(\zeta)] \tag{2}$$

where $\arg[\cdot]$ represents the argument of a complex number.

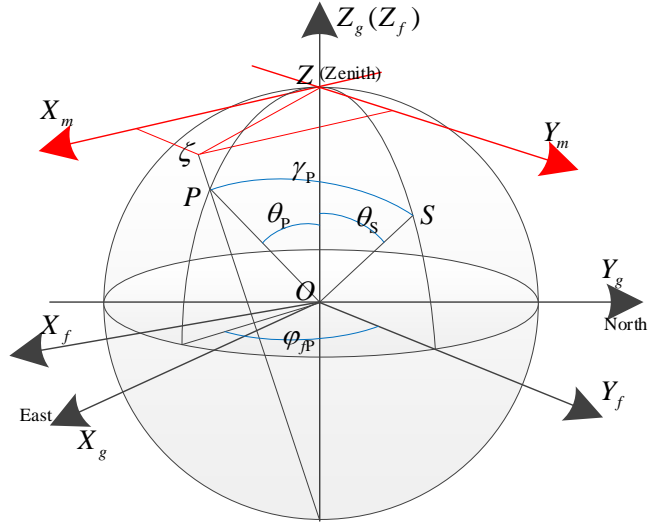

**Fig. 1.** Coordinate systems: $OX_gY_gZ_g$ represents the East-North-Zenith geography coordinate system. $OX_fY_fZ_f$ represents the sun following coordinate, where $Y_f$ is align with the solar azimuth, $Z_f$ points to the sky zenith, $X_f$. $ZX_mY_m$ represents the complex coordinate system, where $Z$ is the sky zenith, $X_m$ is parallel to $X_f$ and $Y_m$ is parallel to $Y_f$. S represents the Sun, $P$ represents the observed point, $\zeta$ represents the projection of $P$ on the complex plane $ZX_mY_m$. $\theta_S$ is the zenith angle of S. $\theta_P$ is the zenith angle of $P$. $\gamma_P$ is the angle between $P$ and S. $\varphi_{fP}$ is the azimuth angles of $P$ in the solar azimuth coordinate system.

DOP:

To well describe the DOP of skylight, the influences of ground albedo, atmospheric turbidity, wavelength are considered [49, 51].

$$DOP(\zeta,\theta_P,\gamma_P,T,\rho,\lambda) = |w(\zeta)| \times (\theta_P E(\theta_P,\rho) + (\frac{\pi}{2}-\theta_P)S(\gamma_P,\theta_P,\lambda))M_{DOP}(T) \qquad (3)$$

where the subscript P represents the observed point. $\theta_P$ represents the zenith angle of P, $\gamma_P$ represents the angle between P and Sun. where $E(\theta_P,\rho)$ represents the depolarization of the earth ground, and $\rho$ represents the ground albedo. $M_{DOP}(T)$ represents the maximal DOP of skylight, and $T$ represents the atmospheric turbidity. $S(\theta_P,\gamma_P,\lambda)$ represents the influence of spectral radiant power, and $\lambda$ represents the wavelength of light.

LI:

To describe the LI of skylight, Hosek model has been proposed to describe the luminance of skylight, which can well describe the measured irradiance curves of skylight [52-54]. Then, the LI of skylight is given by

$$LI(\theta_P,\gamma_P) = (1+Ae^{B/(\cos\theta_P+0.01)}) \cdot (C+De^{E\gamma_P}+F\cos^2\gamma_P+G\cdot\chi(H,\gamma_P)+I\cdot\cos^{1/2}\theta_P) \qquad (4)$$

where $A$, $B$, $C$, $D$, $E$, $F$, $G$, $H$ and $I$ represent adjustable coefficients, which are influenced by ground albedo, atmospheric turbidity and wavelength. $\chi(H,\gamma_P)$ represents the anisotropic term of luminance peaks around the Sun.

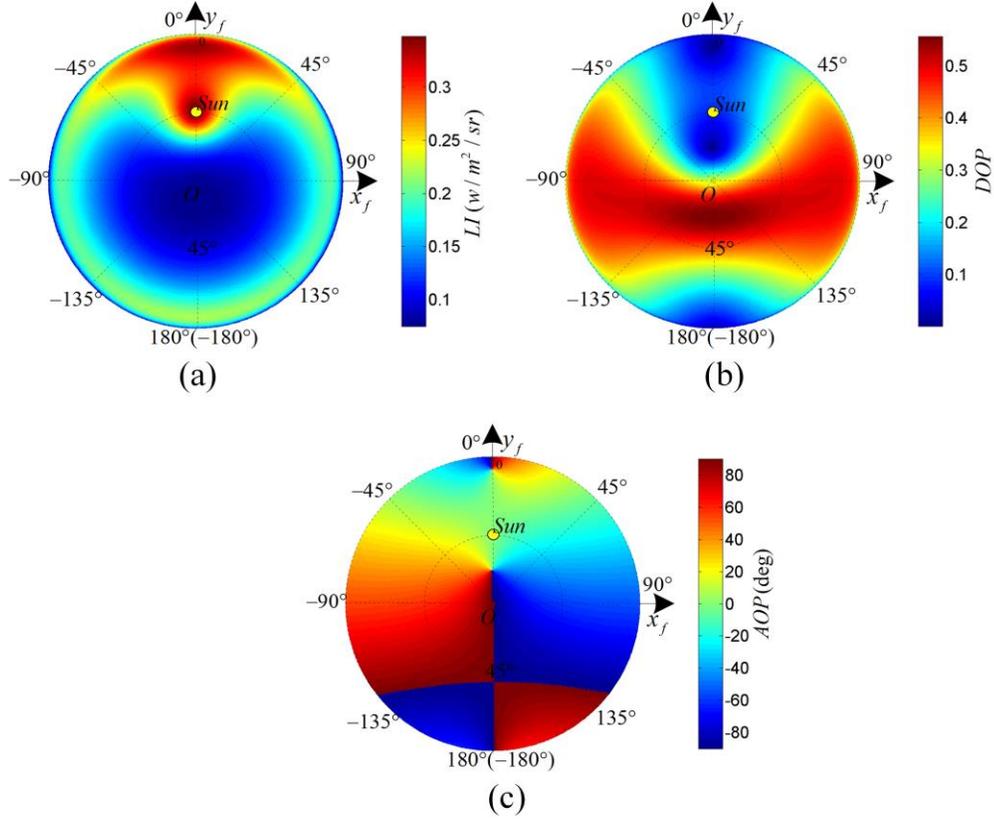

**Fig. 2.** Atmospheric polarization model where, $\theta_S = 45°$, $T = 4$, $\lambda = 450nm$, $\rho = 0.1$: (a) Light intensity (LI). (b) Degree of polarization (DOP). (c) Angle of polarization (AOP).

Above all, a sky model is established. The inputs of this model are the position of the Sun, atmospheric turbidity, wavelength, ground albedo, and position of the observed point. The outputs of this model are DOP and AOP and LI. To facilitate the presentation and discussion, we represent the established sky model in the sun following coordinate $OX_fY_fZ_f$, as shown in Fig. 2.

In addition, to construct a hypothetical polarization imager, the polarization E-vector of the established sky model in sun following coordinate is given by

$$\overrightarrow{E_{fP}} = \left(\sin AOP, \cos AOP, -(\sin AOP \sin \varphi_{fP} + \cos AOP \cos \varphi_{fP}) \tan \theta_P \right)^T \tag{5}$$

where $\varphi_{fP}$ is the azimuth angles of the observed point $P$ in the solar azimuth coordinate system. Then, according to our previous work about hypothetical polarization imager [1], a hypothetical polarization imager of the established sky model in sun following coordinate is constructed, which can capture AOP, DOP and LI images.

## 3. 3D attitude estimation algorithm

To explore whether the skylight polarization patterns contain the information of 3D attitude, a 3D attitude estimation algorithm in terms of SSO [42-44] is proposed in this section. The SSO attitude algorithm considers the difference of each pixel among polarization images based on TM to make full use of the polarization information captured by polarization imager. In the SSO attitude algorithm, the given AOP and DOP images are matched with AOP and DOP images captured by the hypothetical polarization imager of the established sky model. Then, the LI image is used to eliminate invalid solutions.

The SSO algorithm is based upon the simulation of the behavior of social-spiders which considers two different spiders: females and males. Every spider is conducted by many different evolutionary operators depending on gender. Therefore, the SSO algorithm provides excellent performance for TM and reduces computation in the 3D attitude estimation process. The SSO algorithm transforms 3D attitude estimation into an optimization problem. When we estimate 3D attitude, the SSO algorithm is to find the best position of spiders. Every 3D position of spiders corresponds to a 3D attitude. The movement and mating process of male and female spiders will assign new and better 3D positions to spiders, and the iterative process of moving and mating is to solve the best 3D position of spiders. Finally, the best 3D position is obtained, so 3D attitude is estimated. The specific steps of the SSO attitude algorithm to estimate 3D attitude are as follows.

### 3.1 Initializing the population

The numbers of male and female spiders are given based on highly female-biased populations [55]. Suppose the whole population is $N$. Then, the number of females $N_f$ is

$$N_f = floor[(0.9 - rand \cdot 0.25) \cdot N] \tag{6}$$

where $floor(\cdot)$ converts a real number to an integer number, and $rand$ represents a random number between 0 and 1. The number of male spiders $N_m$ is

$$N_m = N - N_f \tag{7}$$

### 3.2 Initializing the positions

The entire spider (female and male) positions are randomly initialized by

$$f_{i,j}^0 = p_j^{low} + rand \cdot (p_j^{high} - p_j^{low}) \quad i=1,2,\cdots,N_f; j=1,2,\cdots,n \tag{8}$$

$$m_{i,j}^0 = p_j^{low} + rand \cdot (p_j^{high} - p_j^{low}) \quad i=1,2,\cdots,N_m; j=1,2,\cdots,n \tag{9}$$

where $i$ is spider indexes. The superscript zero represents the initial population. $n$ represents the dimension of the problem. $m_{i,j}$ represents the $j$th parameter of the $i$th male spider and $f_{i,j}$ represents the $j$th parameter of the $i$th female individual. $p_j^{low}$ is lower initial parameter bound and $p_j^{high}$ is the upper initial parameter bound. For the 3D attitude algorithm, $n=3$. $j=1$ corresponds to the yaw angle $\psi$, $j=2$ corresponds to the pitch angle $\alpha$ and $j=3$ corresponds to the roll angle $\beta$.

For polarization attitude estimation, suppose the field of view of the camera is always above the horizon. So

$$\psi \in [-180°, 180°] \tag{10}$$

$$\alpha \in [-(90 - FOV/2)°, (90 - FOV/2)°] \tag{11}$$

$$\beta \in [-(90 - FOV/2)°, (90 - FOV/2)°] \tag{12}$$

$$|\alpha| + |\beta| \le (90 - FOV/2)° \tag{13}$$

where FOV represents the polarization imager angle of view. Then, we have

$$p_1^{low} = -180° \tag{14}$$

$$p_1^{high} = 180° \tag{15}$$

$$p_2^{low} = -(90 - FOV/2)° \tag{16}$$

$$p_2^{high} = (90 - FOV/2)° \tag{17}$$

$$p_3^{low} = -(90 - FOV/2)° \tag{18}$$

$$p_3^{high} = (90 - FOV/2)°\tag{19}$$

$$|f_{i,2}| + |f_{i,3}| \leq (90 - FOV/2)°\tag{20}$$

$$|m_{i,2}| + |m_{i,3}| \leq (90 - FOV/2)°\tag{21}$$

Because of the symmetry of skylight polarization patterns, TM may appear more than one minimum value. To avoid false matching, we divide the match range into four parts. $(p_1^{low}, p_1^{high})$ is divided into $-180° \sim -90°$, $-90° \sim 0°$, $0° \sim 90°$, and $90° \sim 180°$. Then, we get four minimum matching results. Three of them are excluded by using LI information to obtain the result. The application of LI information will be described in section 3.8.

### 3.3 Fitness assignment

In biology, the characteristic of the spider weight describes the spider capacity. In the SSO algorithm, every spider has a weight $w_i$ which evaluates the solution quality of the individual $i$.

$$w_i = \frac{J(S_i) - worst}{best - worst}\tag{22}$$

where $J(S_i)$ is the weight which is the evaluation of the spider position $S_i$ in terms of the objective function $J(\cdot)$. A larger value of $J(\cdot)$ indicates a better result of TM. The value *best* is the maximum value of $J(S_i)$ and the value *worst* is the minimum. For 3D attitude algorithm based on TM, $J(\cdot)$ is defined as follows

$$J(\cdot) = -\sum_{j_y=-\frac{\eta_y}{2}}^{\eta_y/2} \sum_{i_x=-\frac{\eta_x}{2}}^{\eta_x/2} |DOP_{i_x,j_y}^r - DOP_{i_x,j_y}^s| - \vartheta \sum_{j_y=-\frac{\eta_y}{2}}^{\eta_y/2} \sum_{i_x=-\frac{\eta_x}{2}}^{\eta_x/2} \xi_1(AOP_{i_x,j_y}^r - AOP_{i_x,j_y}^s)\tag{23}$$

where $DOP_{i_x,j_y}^r$ is the DOP of the pixel $(i_x, j_y)$ in the given image. $DOP_{i_x,j_y}^s$ is the DOP of the pixel $(i_x, j_y)$ in the simulation image. $AOP_{i_x,j_y}^r$ is the AOP of the pixel $(i_x, j_y)$ in the given image. $AOP_{i_x,j_y}^s$ is the AOP of the pixel $(i_x, j_y)$ in the simulation image. $\eta_x \times \eta_y$ represent the resolution of polarization imager, $\eta_x$ is the number of pixels in row direction and $\eta_y$ is number of pixels in column direction. $\vartheta$ is a positive proportional coefficient. The definition of $\xi_1(\cdot)$ is shown in Eq. (24), which considers the cyclicity of the AOP ($90°$ is equivalent to $-90°$).

$$\xi_1(x) = \begin{cases} 180 + x & -180° \leq x < -90° \\ -x & -90° \leq x < 0° \\ x & 0° \leq x \leq 90° \\ 180 - x & 90° < x \leq 180° \end{cases}\tag{24}$$

### 3.4 Vibrations through the web

All individuals in the population transmit information mainly by small vibrations in the communal web. The small vibrations depend on the distance and weight of the individual which generates them. The spiders with higher weight generate stronger vibrations. And the spiders receive stronger vibrations compared with spiders located in distant positions. To emulate this fact, the vibrations received by the spider $i$ generated by the spider $j$ are modeled by

$$Vib_{i,j} = w_j \cdot e^{-[\xi_2(d_{i,j})]^2}\tag{25}$$

where $d_{i,j}$ represents the angular distance between the individuals $i$ and $j$. The definition of $\xi_2(\cdot)$ is shown in Eq. (26), which considers the cyclicity of yaw angle ($180°$ is equivalent to $-180°$).

$$\xi_2(x) = \begin{cases} x+360 & -360° \leq x < -180° \\ -x & -180° \leq x < 0° \\ x & 0° \leq x \leq 180° \\ 360-x & 180° < x \leq 360° \end{cases} \quad (26)$$

It is possible to consider vibrations by calculating any pair of spiders, but three special vibrations are calculated within the SSO algorithm:

I. $c$ being the nearest spider to $i$ and having a higher weight compared with $i$ ($Vib_{i,c}$).

II. $b$ being the spider possessing the best weight of the entire population ($Vib_{i,b}$).

III. $f$ being the nearest female spider to $i$ ($Vib_{i,f}$).

### 3.5 Female cooperative operator

The attraction or repulsion between female spiders depends on several random phenomena. To emulate this process, a random number $r_m$ is defined between 0 and 1. If $r_m$ is larger than a threshold $PF$, a repulsion movement is generated. Otherwise, an attraction movement is produced. The female spider position is updated by

$$f_i^{k+1} = \begin{cases} f_i^k + \varsigma \cdot Vib_{i,c} \cdot \xi_2(S_c - f_i^k) + \tau \cdot Vib_{i,b} \cdot \xi_2(S_b - f_i^k) + \delta \cdot (rand - \frac{1}{2}) & r_m < PF \\ f_i^k - \varsigma \cdot Vib_{i,c} \cdot \xi_2(S_c - f_i^k) - \tau \cdot Vib_{i,b} \cdot \xi_2(S_b - f_i^k) + \delta \cdot (rand - \frac{1}{2}) & r_m \geq PF \end{cases} \quad (27)$$

where $\varsigma$, $\tau$ and $\delta$ are random numbers between 0 and 1. $k$ is the iteration number. The spider $S_b$ represents the best spider of the whole population $N$. The spider $S_c$ represents the nearest spider to $i$ that holds a higher weight.

### 3.6 Male cooperative operator

In the biological metaphor, male spiders are divided into two classes: dominant and non-dominant male spiders. Dominant male individuals have better fitness characteristics and are attracted to the closest female individual. On the contrary, non-dominant male individuals concentrate on the center of the male individuals to make the best use of resources that are wasted by dominant male spiders.

$$m_i^{k+1} = \begin{cases} m_i^k + \varsigma \cdot Vib_{i,f} \cdot \xi_2(S_f - m_i^k) + \delta \cdot (rand - \frac{1}{2}) & w_{N_f+i} \geq w_{N_f+m} \\ m_i^k + \varsigma \cdot (\dfrac{\sum_{h=1}^{N_m} m_h^k \cdot w_{N_f+h}}{\sum_{h=1}^{N_m} w_{N_f+h}} - m_i^k) & w_{N_f+i} < w_{N_f+m} \end{cases} \quad (28)$$

where the spider $S_f$ is the nearest female spider to the male spider $i$, $w_{N_f+h}$ represents the weight of a male spider, and $\sum_{h=1}^{N_m} m_h^k \cdot w_{N_f+h} / \sum_{h=1}^{N_m} w_{N_f+h}$ represents the weighted mean of the male spiders.

### 3.7 Mating operator

When a dominant male individual locates some female spiders within a mating range $r$, it mates and generates a new spider to replace the worst spider. In the mating process, the new brood is assigned by the roulette method [42]. For the 3D attitude algorithm, according to Eq. (14)- (21). radius $r$ is computed according to the average of parameter bounds.

$$r = \frac{\sum_{j=1}^{n}\left(p_j^{high} - p_j^{low}\right)}{2n} \quad (29)$$

*3.8 Final solution determination*

The LI value obtained by the polarization imager is relative values and susceptible to interference. So the LI images cannot be directly used to TM. But the LI images can be used to determine the final solution. Remove two more distant solutions according to the TM distance, then the final solution is determined in terms of LI images. As shown in Fig. 3, the LI image is divided into 8 regions and the proportions of total LI in 8 regions are obtained. The proportions of the given LI image are compared with that of simulation LI images using Eq. (30).

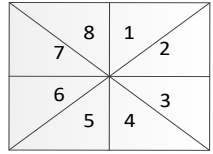

**Fig. 3.** Eight regions of the LI image.

$$Com = \sum_{n=1}^{4}(LI_n^r / LI_{n+4}^r - LI_n^s / LI_{n+4}^s)^2 \quad n=1,2,3,4 \quad (30)$$

where $LI_n^r$ is the LI of the region $n$ in given image. $LI_n^s$ is the LI of the region $n$ in simulation image. And the result closed to the given LI image is considered to be the final solution.

*3.9 Computational Procedure*

The steps of the 3D attitude algorithm are shown as follows:
    Step 1: Initialize the number of spiders (female and male) (section3.1).
    Step 2: Initialize the entire spider positions (section 3.2).
    Step 3: Calculate the weight of every spider in terms of TM (section 3.3).
    Step 4: Calculate the vibrations through the communal web (section 3.4).
    Step 5: Move the female spiders (section 3.5).
    Step 6: Move the male spiders (section 3.6).
    Step 7: Perform the mating operation (Section 3.7).
    Step 8: If the stop criteria are met, go to step 9; Otherwise, go back to Step 3.
    Step 9: Consider the LI images to obtain the final solution (section 3.8).

## 4. Simulations

Based on the established sky model and SSO attitude algorithm, simulations are carried out to explore whether the established sky model contains the information of 3D attitude and the influence of measurement noise and model error on 3D attitude estimation. The issues, which we are particularly interested in, are
    1. 3D attitude information exploration.
    2. Sensitivity to measurement error.
    3. Sensitivity to model error. We want to investigate how slight errors in atmospheric turbidity $T$ and ground albedo $\rho$ affect the performance of the 3D attitude algorithm. The

influence of wavelength $\lambda$ is determined by the filter in front of the polarization imager, so its impacts don't be considered here.

Table 1 shows the parameters used in simulations.

For the actual atmosphere, the value of atmospheric turbidity is usually between 3 and 7 [56], the value of ground albedo is usually between 0.1 and 0.4 [57]. The wavelength range is between 320nm and 720nm, which depends on the actual polarization imager. These three parameters are mean distribution within the given ranges. What needs to be specially explained here is the influence of the solar altitude angle $h_s$ ( $h_s$=90°-$\theta_s$ ). When the sun is at the sky zenith point, the orientation cannot be estimate, therefore we study the relation between the errors of 3D attitude estimation and solar altitude angle.

**Table 1. Simulation Parameters**

| Quantity | Value | Units | Description |
| --- | --- | --- | --- |
| $N$ | 200 | / | Entire spider population |
| $k$ | 1000 | / | Iteration number (Stop criteria) |
| $PF$ | 0.7 | / | Threshold |
| $\vartheta$ | 1.5 | / | Proportional coefficient |
| $FOV$ | 107.95 | deg | Angle of view |
| $\rho$ | 0.1~0.4 | / | Ground albedo |
| $\lambda$ | 320~720 | nm | Wavelength |
| $h_s$ | 0~90 | deg | Solar altitude angle |
| $\eta_x$ | 2048 | pixel | Number of pixels in row direction |
| $\eta_y$ | 2448 | pixel | Number of pixels in column direction |

*4.1 3D attitude information exploration*

The 3D attitude information of the established sky model is explored, with the polarization imager noise set to 0. Besides, $T$ and $\rho$ of the established sky model, which are used to capture polarization imagers, are exactly equal to that used to capture the given polarization images.

As shown in Fig. 4, the maximum errors (MaxE) of pitch and roll angles are all less than $0.38°$ under perfect date. Moreover, mean absolute errors (MAE) and root mean square errors (RMSE) of pitch and roll angles are all less than $0.018°$ and $0.036°$ for any solar altitude angle. When the solar altitude angle is not $90°$, the MAE, RMSE and MaxE of yaw angle estimation is always less than $0.015°$, $0.024°$, $0.34°$ respectively. But when the solar altitude is $90°$, the MaxE of yaw angle estimation reaches $180°$. This is due to, when the solar altitude angle is $90°$, two neutral points (Babinet and Brewster neutral points) coincide to become a neutral point [58, 59] and the projection of the sun vector on the horizontal plane is 0 vector. These leads to the yaw angle cannot be estimated when solar altitude is $90°$. So, the solar altitude angle is usually limited to $0° \sim 60°$ when using polarized light to estimate yaw angle [25].

So when solar altitude angle is between $0°$ and $60°$, the simulation results of 3D attitude information exploration show that 3D attitude can be estimated by the designed SSO algorithm under perfect date, the accuracy of 3D attitude estimation meet the basic requirements of navigation. And, the results demonstrate that the established sky model contains 3D attitude information under the common meteorological conditions. The actual skylight polarization patterns, which are often affected by clouds, dust and so on, are more complex. Besides, only one wavelength of light is used in this simulation, multispectral information can also be used in practice 3D attitude estimation. So, it can be inferred that the actual skylight polarization patterns are very likely to contain 3D attitude information under the common meteorological conditions.

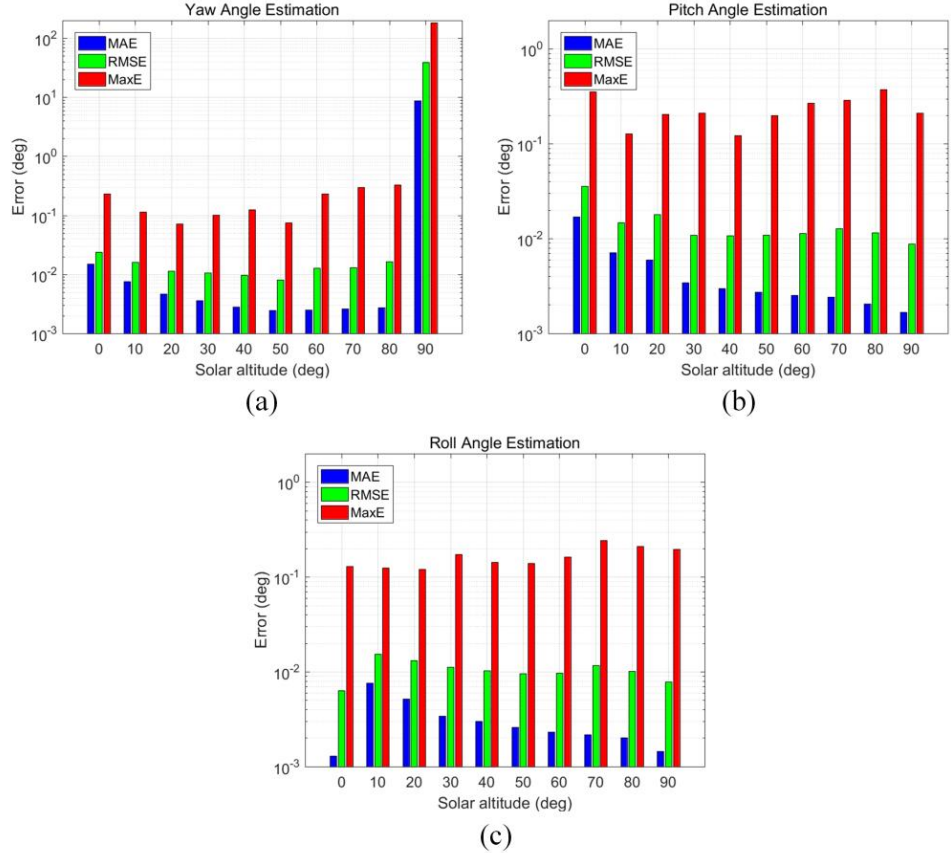

**Fig. 4.** Results of 3D attitude information exploration. (a) Yaw angle error. (b) Pitch angle error. (c) Roll angle error.

### 4.2 Sensitivity to measurement error

To explore the influence of measurement error on 3D attitude estimation, Gaussian noise is added to the given polarization images. The mean of Gaussian noise is 0, and the variance of Gaussian white noise is 5%(maximum pixel value minus minimum pixel value), in other words, the mean and variance of Gaussian white noise added to DOP image are 0 and 5%(maximum DOP minus minimum DOP), the mean and variance of Gaussian white noise added to AOP image are 0 and 5%(maximum AOP minus minimum AOP), the mean and variance of Gaussian white noise added to LI image are 0 and 5%(maximum LI minus minimum LI).

As shown in Fig. 5, the MaxE, MAE and RMSE of three Euler angles are all greater than 0.72°, 0.013° and 0.070° under measurement noise. It can be found that when the solar altitude angle is 90°, the MaxE of yaw angle estimation is 180°. This is due to two neutral points coincide to become a neutral point and the projection of the sun vector onto the horizontal plane is the 0 vector, as mentioned in section 4.1. Moreover, when the solar altitude angle is 0°, the MaxE of yaw angle estimation also reaches 180°. This is probably due to the symmetry of the sky model. For Rayleigh sky model, because of its plane symmetry about the plane 90° away from the sun, the sun vector obtained from Rayleigh sky model is a bidirectional vector [1], which lead to the range of yaw angle estimated by polarized skylight is $0° \sim 180°$. If the range of yaw angle estimated by polarized skylight is expected to be $0° \sim 360°$, additional determination conditions are required. The designed sky model in this paper is not strictly symmetry about the plane 90° away from the sun. So, without the influence of noise, the yaw angle can be estimated in the range $0° \sim 360°$. However, the designed sky model has a certain symmetry about the plane 90° away from the sun, especially when the solar altitude angle is 0°, this plane symmetry is more obvious. So, when there are noise and the solar altitude angle is 0°, it is difficult to estimate the yaw angle in the range of $0° \sim 360°$, so yaw angle has the errors of about 180°.

Above all, the results of sensitivity to measurement error are compared with that of 3D attitude information exploration, we can find that the errors of three Euler angles all increase a lot, so 3D attitude estimation is sensitive to measurement error.

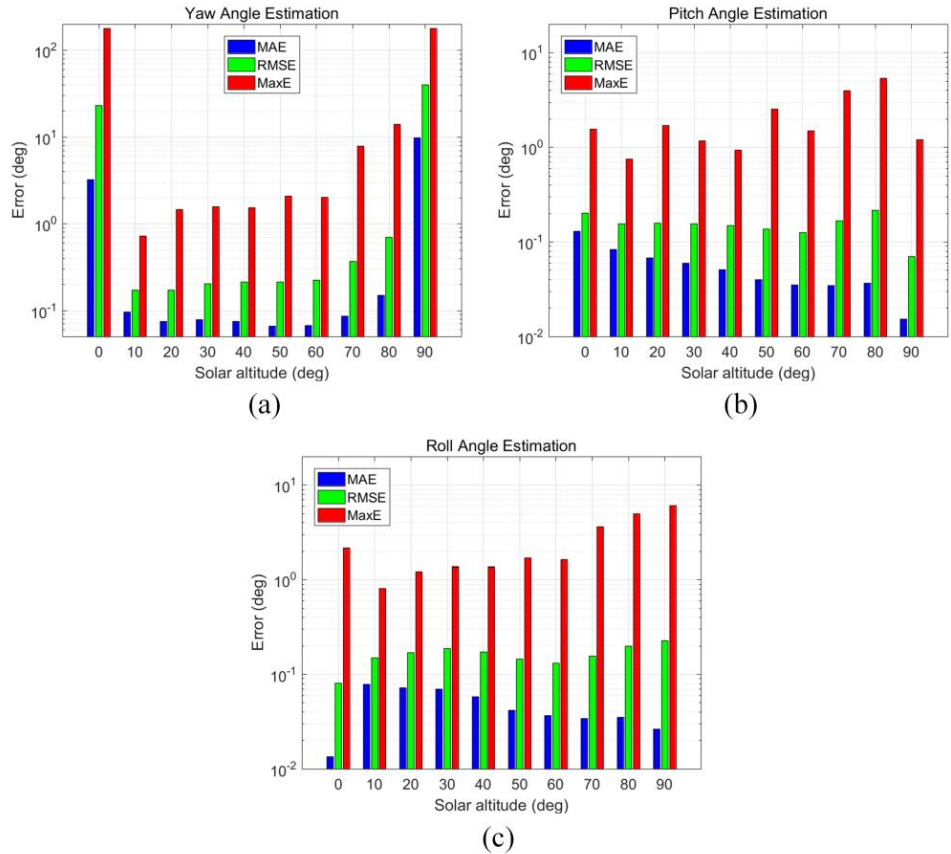

**Fig. 5.** Results of sensitivity to measurement error. (a) Yaw angle error. (b) Pitch angle error. (c) Roll angle error.

## 4.3 Sensitivity to model error

Atmospheric turbidity $T$ and ground albedo $\rho$ are hard to predict and sometimes impractical to measure accurately. So we want to investigate how slight errors in atmospheric turbidity $T$ and ground albedo $\rho$ affect 3D attitude estimation. In this section, the performance of 3D attitude estimation is investigated when $T$ error is 5% and $\rho$ error is 5%.

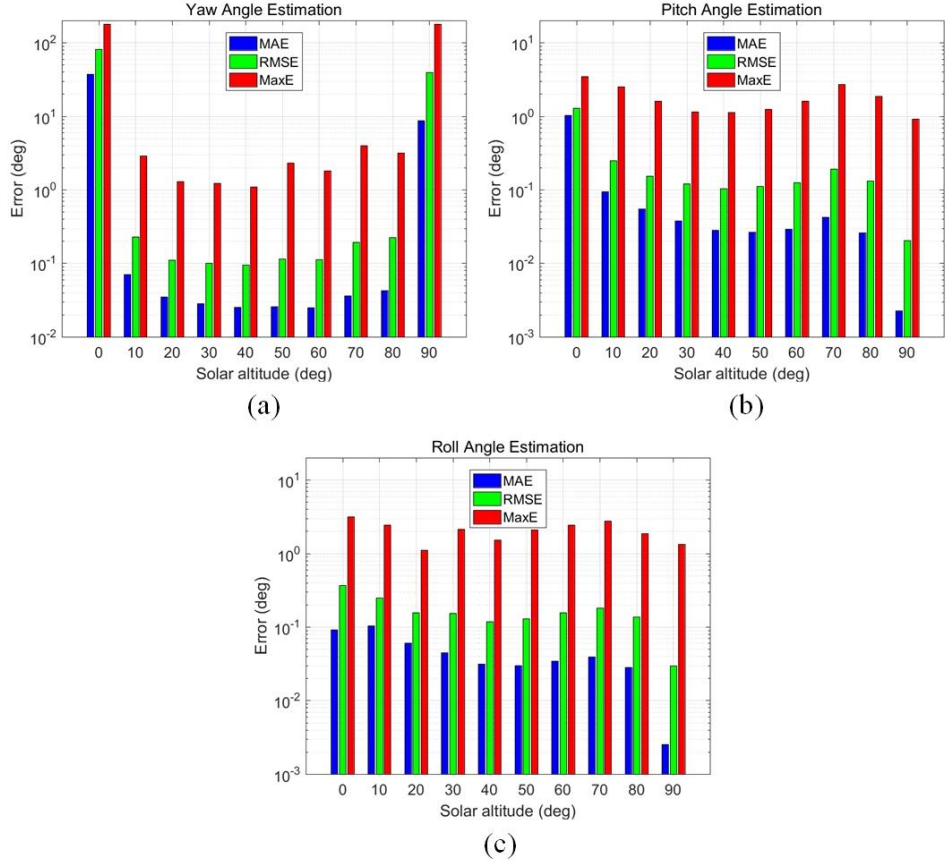

**Fig. 6.** Results of sensitivity of model error. (a) Yaw angle error. (b) Pitch angle error. (c) Roll angle error.

Simulation results are represented in Fig. 6. Under model error, the MaxE, MAE and RMSE of three Euler angles are all greater than 0.91°, 0.002° and 0.020° respectively. In addition, when the solar altitude angle is 90°, the MaxE of yaw angle estimation is 180°. This is due to two neutral points coincide to become a neutral point and the projection of the sun vector onto the horizontal plane is the 0 vector, as mentioned in section 4.1. When the solar altitude angle is 0°, the MaxE of yaw angle estimation also reaches 180°. This is probably due to the certain symmetry of the sky model, as mentioned in section 4.2.

The results of sensitivity to model error are compared with that of 3D attitude information exploration, we can find that the errors of three Euler angles increase a lot. Therefore, the accuracy of the sky model is very important for 3D attitude estimation.

## 5. Experiment

For actual polarized skylight navigation, measurement error has not only Gaussian white noise, but also measurement systematic error, gross error and so on; Model errors are also more complex. So, it is very difficult to obtain 3D attitude only by actual polarized skylight.

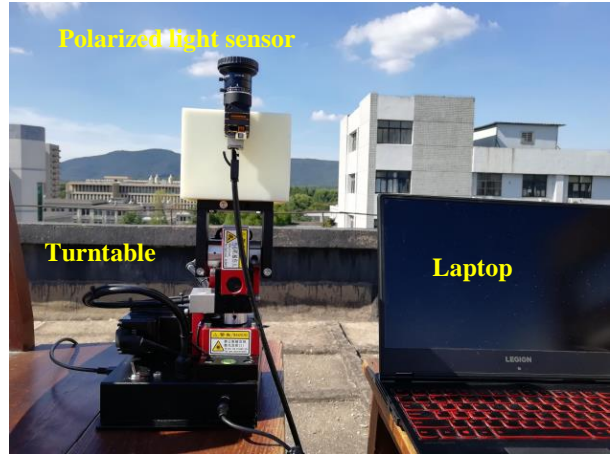

**Fig. 7.** Experiment platform.

Even so, a simple experiment was carried out in the campus of Nanjing University of Science and Technology (32°01′36.5″N, 118°51′11.8″E), and the ground albedo of experiment site is 0.16. the experiment platform is shown in Fig. 7. The polarized light sensor is Sony phoenix PHX050S-P, whose parameters are same as the hypothetical polarization imager in simulation. This sensor is installed on a turntable, whose angle resolution is $0.06°$. The experiment was carried out on August 15, 2020, from 8:11 to 18:40, the weather was clear and there were a few clouds in the sky, and the atmospheric turbidity is 3.6~5.5.

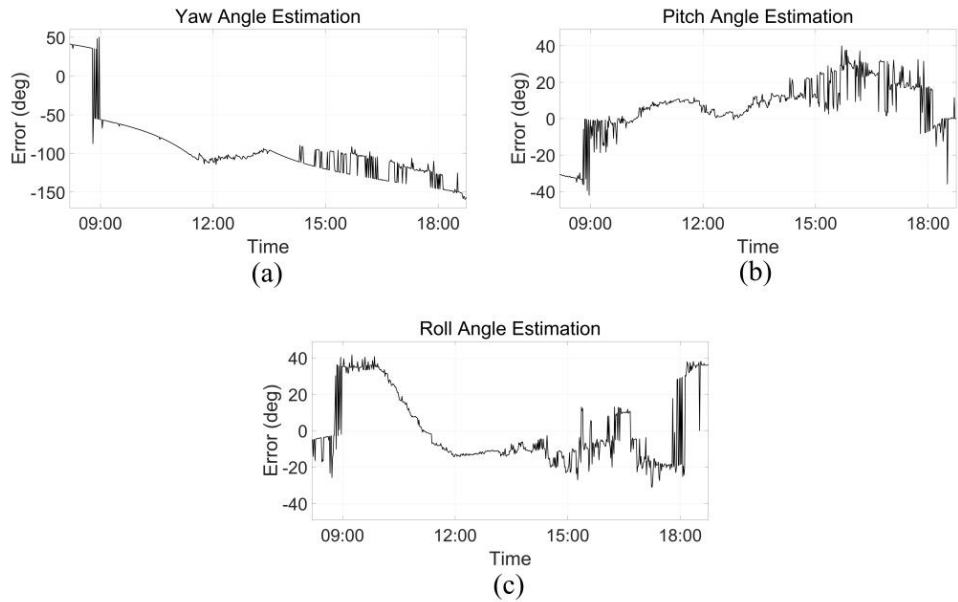

**Fig. 8.** Results of experiment. (a) Yaw angle error. (b) Pitch angle error. (c) Roll angle error.

The experiment results are shown in Fig. 8. It can be found that the errors of 3D attitude estimation are very large and unable to meet the basic needs of navigation. This is due to that 3D attitude estimation requires much higher accuracy of sky model, much higher precision of polarized light sensor, and much more stable algorithm than 1D and 2D attitude estimation. The measurement noise of the actual polarization sensor may have not only Gaussian white noise, but also measurement systematic error, gross error and so on. The sky model is also affected by many meteorological factors, and the actual model error is more complex. It can be expected that it is very difficult to estimated 3D attitude only by polarized light under the limited condition of sensor accuracy, model accuracy and algorithm stability.

## 6. Discussion and conclusion

Based on the established sky model and SSO attitude algorithm, the simulation results demonstrate that the designed SSO algorithm can estimate 3D attitude and the established sky model contains 3D attitude information under perfect date and normal weather conditions. The actual skylight polarization patterns are more complex and influenced by more factors. In addition, only one wavelength of light is used for 3D attitude estimation in our simulation, however, multispectral information can also be used in practice 3D attitude estimation. So, it can be inferred that the actual skylight polarization patterns are very likely to contain 3D attitude information under the common meteorological conditions. This provides an important reference for the future application of bioinspired polarized skylight navigation on 3D attitude estimation.

However, the simulations of sensitivity to measurement error and model error demonstrate that the slight measurement noise and model error will seriously affect the accuracy of 3D attitude estimation. The experimental results also show that it is very difficult to estimate 3D attitude based only on skylight polarization patterns. This is due to that 3D attitude estimation requires much higher accuracy of sky model, much higher precision of polarized light sensor, and much more stable algorithm than 1D and 2D attitude estimation. There are very few feature points available in the skylight polarization pattens. Rayleigh sky model only have two feature points (Sun and anti-Sun points). So, Rayleigh sky model does not contain 3D attitude information. The established sky model has four feature points (four neutral points) and considers some other factors. So, it contains 3D attitude information. But the feature points of this model are still very few, so that the slight measurement noise and model error will seriously affect the accuracy of 3D attitude estimation. Therefore, when only using polarized skylight to estimate three Euler angles, an accurate skylight polarization model with more feature points and high precision polarized light sensors are needed.

A 3D attitude algorithm based on SSO is proposed in this paper, which can make full use of the collected polarization information and estimate three Euler angles under perfect date and normal weather conditions. However, this algorithm takes a long time because it needs many iterations and the spider positions of this algorithm are generated randomly, which leads to certain randomness of the results of attitude estimation. And because of the difficulty of obtaining 3D attitude only by polarized skylight, the SSO algorithm and sky model designed in this paper have no practical application value at present. All these show that it is very difficult to determine 3D attitude in real time based only on polarized skylight patterns. But this problem very interesting and worth exploring and studying, and we will continue to conduct relevant research in the future.

## Funding